# Cascade-Forward Neural Network Based on Resilient Backpropagation for Simultaneous Parameters and State Space Estimations of Brushed DC Machines

Hacene MELLAH[1,2*], Kamel Eddine HEMSAS[2], Rachid TALEB[3]

[1] Electrical engineering department, faculty of sciences and applied sciences, University of Akli Mouhand Oulhadj-Bouira, Algeria
[2] Department of Electrical Engineering, Faculty of Technology, Ferhat Abbas Sétif 1 University, LAS laboratory, Sétif, Algeria
[3] Department of Electrical Engineering, LGEER Laboratory, Hassiba Benbouali University, Chlef, Algeria

Corresponding Author Email: has.mel@gmail.com

**ABSTRACT**

A sensorless speed, average temperature and resistance estimation technique based on Neural Network (NN) for brushed DC machines is proposed in this paper. The literature on parameters and state spaces estimations of the Brushed DC machines, shows a variety of approaches. However, these observers are sensitive to a noise, on the model accuracy also are difficult to stabilize and to converge. Furthermore, the majority of earlier works, estimate either the speed or the temperature or the winding resistance. According to the literatures, the Resilient backpropagation (RBP) as is the known as the faster BP algorithm, Cascade-Forward Neural Network (CFNN), is known as the among accelerated learning backpropagation algorithms, that's why where it is found in several researches, also in several applications in these few years. The main objective of this paper is to introduce an intelligent sensor based on resilient BP to estimate simultaneously the speed, armature temperature and resistance of brushed DC machines only from the measured current and voltage. A comparison between the obtained results and the results of traditional estimator has been made to prove the ability of the proposed method. This method can be embedded in thermal monitoring systems, in high performance motor drives.

## 1. INTRODUCTION

In the last few years there has been a growing interest in thermal aspects of electrical machines and their effects, the electrical and mechanical time constants varied for each temperature variation, also the electrical resistance and its back EMF depend on temperature [1]; during operation, the characteristics, performance of electric motors were not the same as those the design's [2], as a result, the temperature quantification is very important to the best control and the reliability of electrical machines.

The normal effect of thermal aging is to make the insulation system vulnerable to other factors and effects that currently produce failures [3, 4]. Once the insulation loses its physical performance, it can no longer withstand the various dielectric, mechanical and environmental effects, because of these catastrophic effects many researchers interested in the insulation systems monitoring methods of electrical machines [5]. Among the causes of thermal faults are: overloads [6], cyclic mode [7], over voltage and unbalances voltage [8], distortion voltage [4], thermal insulation aging [3], obstructed or impaired cooling [9], poor design and manufacture [3], skin effect [10], the interested reader is referred to [3-10] for more detailed about the cause of stator and rotor failures.

For several years, great effort has been devoted to the temperature and speed measurement of electrical machines, in literature, we find several methods about temperature [11-13] and speed measurements [14] of electrical apparatus. The direct measurement of temperature in electric DC machines is an old theme treated at their time with less pressure [13, 15, 16], on the other hand, (indirect) some author obtained the average winding temperature from the resistance measurement [13], a more modern method can be found in [12, 17, 18], but measurement of the temperature poses two major problems: the measurement point i.e the optimum sensor placement and the obtaining of the thermal information from the rotor [18], in the same manner for the speed measurement, some difficulties are presented her [19].

Moreover, obtaining information from sensors installed on the armature adds techno-economics difficulties on the measurement chain; these technical and economic disadvantages of physical sensors as well known to researchers, pushes them for sensorless solutions [17, 20, 21].

To solve the problem of sensorless speed estimation, many researchers have proposed various methods [22, 23], a position-sensorless control of brushless DC motor for electric vehicles application is presented in [24], a low-cost low-resolution sensorless for brushed DC motor is proposed and experimentally validated in [25], a sensorless estimation based on support vector machines is proposed by [26]. however, [27] suggest a speed estimation based quantized sensors of PMDC motors. An excellent review about position-sensorless operation of brushless permanent-magnet machines is presented in [22].

One of the first examples of temperature estimation is presented in [28], when the authors apply a Luenberger observer both for DC rolling mill motor and a squirrel cage induction motor, another solution is described in [29] where the authors use a steady-state EKF associated with its transient version, nevertheless, for the resistance estimation some author combine between EKF with the smooth variable structure filter [30].

Some research on bi-estimation has been done [31, 32], in our point view the most interesting approach to this issue has been proposed by Acarnley et al. in [32], where they propose, applies and experimentally validated the transient EKF to estimate the speed and armature temperature in a brushed DC motor. However, we can summarize the EKF limitations for three points, if the system is incorrectly modelled the filter could quickly diverge, the EKF assumes that the noises are Gaussian [33-35] may not be the reality [36] and eventually, if the initial state estimate values are incorrect also the filter may diverge [37]. Furthermore, using an EKF, which is difficult to stabilize with the sensible choices of covariance matrices [34-36].

However, to the authors' knowledge, very few publications can be found in the literature dealing with the simultaneous estimation of speed, armature temperature of brushed DC machines [32], especially by intelligent estimators based on NN [38], despite the NN has been applied to process control [39], diagnostics, identification [40], prediction [41], power electronics [42] and robotics [43], social studies [44], building [45] and medical [46].

In the paper [38], the authors discuss how to avoid the limits of the standard NN based on Multilayer Perceptron with Levenberg-Marquardt Backpropagation in their application, and propose as a solution a CFNN based on Bayesian Regulation backpropagation (BRBP). However, an NN based on BRBP is very accurate but need an enormous time to converge and is known as a slow algorithm to converge [44, 45], based on the approach presented in [38], the purpose of this paper is interest to a CFNN based on fast learning algorithm. According to the literatures the Resilient backpropagation (RBP) as is the known as the faster BP [46-50], the main objective of this paper is to introduce an intelligent NN-based resilient BP sensor to estimate simultaneously the speed, armature temperature and resistance only from the measured current and voltage.

The remainder of the paper is organized as follows sections: Section II describes the thermal model of Brushed DC motor, Section III discusses on the NN and CFNN based on RBP and give some detail about RP properties and its variants. Simulation results are presented, commented and compared with the earlier results in Section IV; Section V concludes the paper.

## 2. THERMAL MODEL OF BRUSHED DC MOTOR

The researchers begin to interest to study of rotating electric machinery from the combined viewpoints of thermal and electrical processes from the last middle century [51, 52]. The model used in this paper is proposed by [32], the electrical equation can write as:

$$V_a = R_{a0}(1 + \alpha_{cu}\theta)i_a + l_a \frac{di_a}{dt} + k_e\omega \quad (1)$$

Where: $V_a$ is armature voltage, $R_{a0}$ is armature resistance at ambient temperature, $\alpha_{cu}$ ($\alpha_{cu}$ = 0.004 /°C) temperature coefficient of resistance, $\theta$ temperature above ambient, $i_a$ armature current, $l_a$ is armature inductance, $k_e$ is torque constant, and $\omega$ armature speed. The mechanical equation:

$$J \frac{d\omega}{dt} + b\omega + T_L = k_e i_a \quad (2)$$

where $J$ (kg × m²) is total inertia, $b$ (N × m × s) is the viscous friction constant, and $T_L$ (N × m) is the load torque.

The thermal model is derived by considering the power dissipation and heat transfer [32]. The power dissipated by the armature current flowing through the armature resistance, which varies in proportion to the temperature. The iron loss is proportional to speed squared for constant excitation multiplied by the iron loss constant $k_{ir}$ ($k_{ir}$ = 0.0041 W/(rad/s)²). The power losses $P_{lo}$ include contributions from copper losses and iron losses which frequency dependent:

$$P_{lo} = R_{a0}(1 + \alpha_{cu}\theta)i_a^2 + k_{ir}\omega^2 \quad (3)$$

Heat flow from the DC motor is either directly to the cooling air and depends on the thermal transfer coefficients at zero speed $K$ ($K$ =4.33 W/°C) and with speed $KS$ ($KS$ = 0.0028 s/rad); The thermal power flow from the DC motor surface that is proportional to the difference temperature between the motor and the ambient air temperature, and the temperature variation in the armature which depends on the thermal capacity $H$ ($H$=18 KJ/°C):

$$P_{lo} = K(1 + KS\omega)\theta + H\frac{d\theta}{dt} \quad (4)$$

By arranging the previous eqs, we can write the equations system as:

$$\begin{aligned} \frac{di_a}{dt} &= -\frac{R_{a0}(1+\alpha_{cu}\theta)}{l_a}i_a - \frac{k_e}{l_a}\omega + \frac{1}{l_a}V_a \\ \frac{d\omega}{dt} &= \frac{k_e}{J}i_a - \frac{b}{J}\omega - \frac{1}{J}T_L \\ \frac{d\theta}{dt} &= \frac{R_{a0}(1+\alpha_{cu}\theta)}{H}i_a^2 + \frac{k_{ir}}{H}\omega^2 - \frac{K(1+KS\omega)}{H}\theta \end{aligned} \quad (5)$$

## 3. ANN ESTIMATOR

In recent years, several authors use a cascade forward backpropagation neural network (CFNN) and has become very popular [53-75] a CFNN proved their capability in several applications and it they become their preferred choice [74].

Many authors [70-79], assert that the CFNN are similar to FFNN, but include a weight connection from the input to each layer and from each layer to the successive layers. As example, a four-layer network has connections from layer 1 to layer 2, layer 2 to layer 3, layer 3 to layer 4, layer 1 to layer 3, layer 1 to layer 4 and layer 2 to layer 4. In addition, the four-layer network also has connections from the input to all layers. As FFNN and CFNN can potentially learn any input-output relationship, but the CFNNs with more layers might learn complex relationships more quickly [74-76, 80], which makes it the right choice for intended for accelerated learning in NNs [75]. The results obtained by Filik et al. in [80] suggest that which cascade forward back propagation method can be more effective than feed-forward back propagation method in some cases. And on the other hand, the FFNN cannot solve some problems [77]. the reader is referred to [74-77, 79, 80] for more detailed.

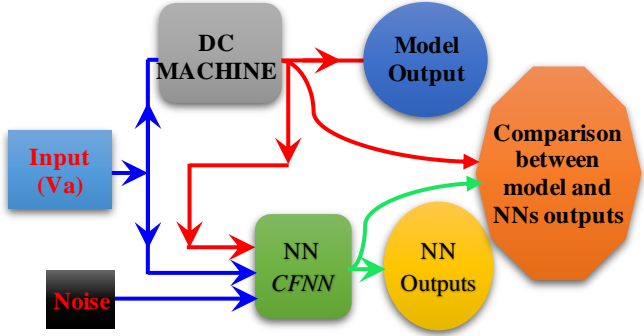

**Figure 1.** Comparison between model and NN's outputs

In this application, the CFNN inputs are the voltage and current and the outputs are the speed and the armature temperature and resistance, to test the robustness and to make the CFNN's inputs similar to the output of the sensor for the real-time applications, a random white Gaussian noise has been added to the inputs patterns.

### 3.1 Back-propagation training algorithms

The backpropagation algorithm is used to form the neural network such that on all training patterns, the sum squared error 'E' between the actual network outputs, 'y' and the corresponding desired outputs, yd, is minimized to a supposed value:

$$E = \sum (y_d - y)^2 \qquad (6)$$

To get the optimal network architecture, for each layer the transfer function types must be determined by trial and error method. On the input and hidden layer, a hyperbolic tangent sigmoid transfer function has been used, defined as:

$$f(net_j) = \frac{2}{1+e^{-2net_j}} - 1 \qquad (7)$$

where net is the weighted sum of the input unit, and f(net) is the output units. For the output layer has 3 units with a pure linear transfer function.

$$f(net_j) = net_j \qquad (8)$$

### 3.2 Principe and rule

Resilient backpropagation often abridged by Rprop [49, 49, 81-84] or RBP [46, 48, 58, 78, 85] was created by M.Riedmiller *et al* in 1992 [81], is a learning heuristic [49] and is a batch update algorithm [86] for supervised learning [84, 87] and Rprop is a first-order optimization algorithm [88].

Rprop performs a local adaptation of the weight-updates based on the sign of the partial derivative ∂E/∂wij to eliminate the harmful influence of the size of the partial derivative on the weight step. It is based on the so-called Manhattan Learning rule [81, 83], for more details the reader is referred to [48, 49, 81-84, 89].

In each iteration, the new weights are given by:

$$w_{ij}^{(t+1)} = w_{ij}^{(t)} + \Delta w_{ij}^{(t)} \qquad (9)$$

The size of the weight change is exclusively determined by a weight-specific, so-called 'update-value' performed as follows:

$$w_{ij}^{(t)} = \begin{cases} -\Delta_{ij}^{(t)}, & \text{if } \dfrac{\partial E^{(t)}}{\partial w_{ij}} > 0 \\ +\Delta_{ij}^{(t)}, & \text{if } \dfrac{\partial E^{(t)}}{\partial w_{ij}} < 0 \\ 0, & \text{otherwise} \end{cases} \qquad (10)$$

The second step of Rprop learning is to determine the new update-values, the step size update rules are:

$$\Delta_{ij}^{(t)} = \begin{cases} \eta^+ \cdot \Delta_{ij}^{(t-1)}, & \text{if } \dfrac{\partial E^{(t-1)}}{\partial w_{ij}} \cdot \dfrac{\partial E^{(t)}}{\partial w_{ij}} > 0 \\ \eta^- \cdot \Delta_{ij}^{(t-1)}, & \text{if } \dfrac{\partial E^{(t-1)}}{\partial w_{ij}} \cdot \dfrac{\partial E^{(t)}}{\partial w_{ij}} < 0 \\ \Delta_{ij}^{(t-1)}, & \text{otherwise} \end{cases} \qquad (12)$$

With $0 < \eta^- < 1 < \eta^+$, For each weight, if there was a sign change of the partial derivative of the total error function for two successive iteration, the update value for that weight is multiplied by a factor $\eta^-$, where $\eta^- < 1$, the preferred value of the decrease factor which gives us the best results is $\eta^-=0.5$ [84, 87], but if two successive iteration produced the same sign, the update value is multiplied by a factor of $\eta^+$, where $\eta^+ > 1$, the preferred value of the increase factor which gives us the best results is $\eta^+=1.2$ [84, 87], the maximum weight step is fixed to $\Delta_{max}=50$, and the minimum step-size is $\Delta_{min}=10^{-6}$ [84, 87], for more detailed the interested reader is referred to [49, 81-84, 87, 89].

### 3.3 Rprop Variants

Two variants have been firstly created, with weight-backtracking [83, 84] named Rprop+ [82] and without weight-backtracking [87] named Rprop− [82]. A performance comparative studies between these algorithms and many other of feedforward supervised learning techniques for many benchmark problems has been presented in [84, 87].

Igel et al create two new versions is based to adding a stored the previous error E(t−1) as a new variable to Rprop+, this version named iRprop+ [82], the second one is that the derivative (∂E(t)/∂wij) is set to zero [82], iRPROP− is described [49, 82], so, the only difference between Rprop− and iRprop− is that the derivative (∂E(t)/∂wij) is set to zero [82], and as comparison between iRprop− and iRprop+, iRprop− is the same as iRprop+, but without weight-backtracking [49]. The reader is referred to [49,81-87] too well understood these variants, where a performances comparison of all Rprop variants and several learning algorithms has been carried out with four neural network benchmark problems.

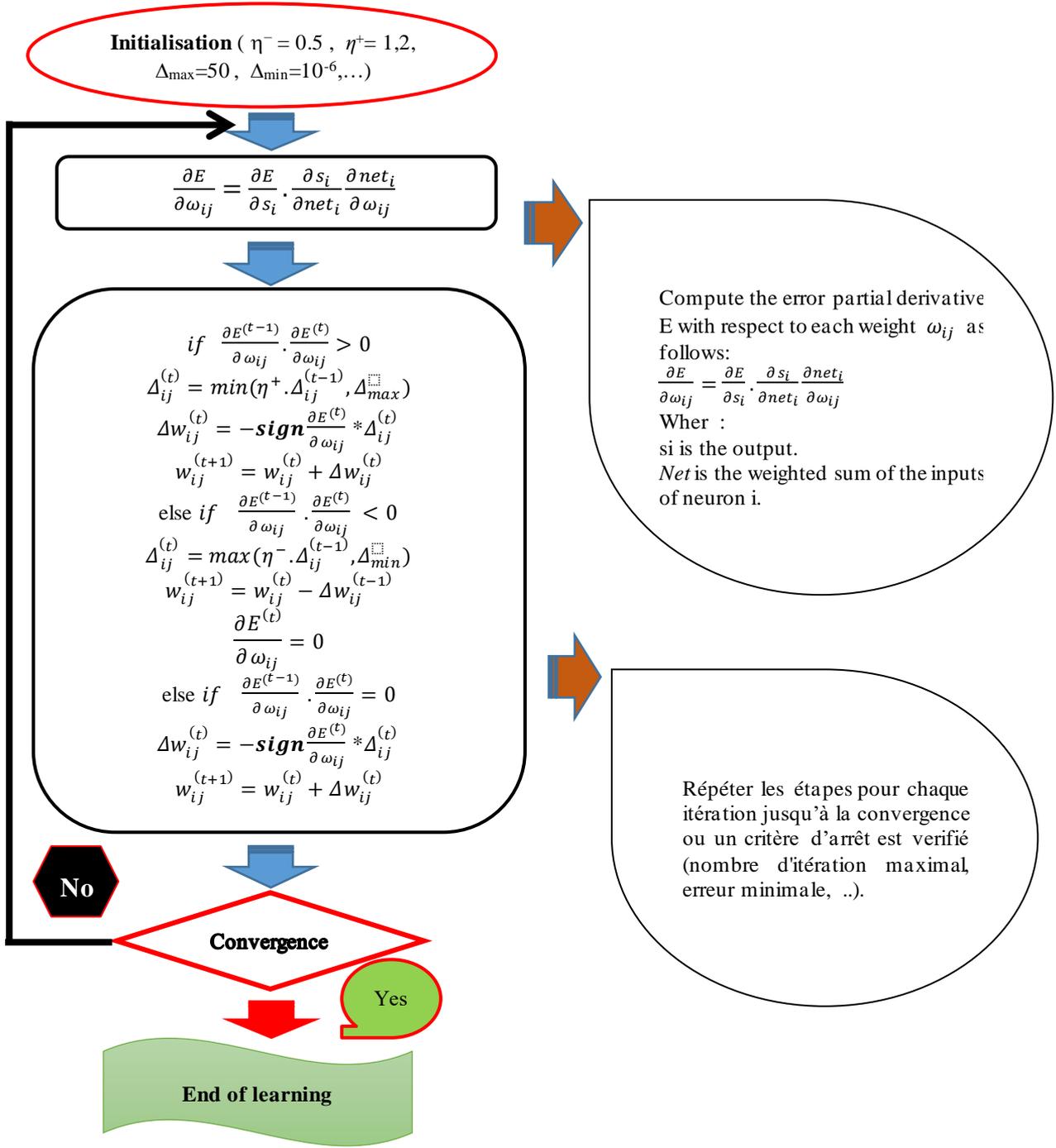

**Figure 2.** Procedures and steps of ANN based on a Resilient backpropagation learning algorithm

## 4. SIMULATION RESULTS

The procedure how the simulation data were used to train the NN is the cross-validation error checked for multiple sets of training data, this data is the result of the equation (4) with the use of the parameters of BDC motor shown in table 1.

The estimated speed, armature temperature and resistance are shown in Figs. 3-6 for a continuous running duty or abbreviated by duty type S1. where duty type S1 characterized by an operation at a constant load maintained for sufficient time to allow the machine to reach thermal equilibrium [90].

*Table 1.* Parameters of BDC motor used in the simulation.

| **Rated voltage** | Va = 240 V |
|---|---|
| **Power** | P = 3 kW |
| **Rated torque** | TL = 11 N.m |
| **Armature resistance** | Ra = 3.5 Ω |
| **Armature inductance** | $l_a$ = 34 mH |

The estimated speed and the corresponding errors are shown in Figure 3, the results obtained by Acarnley et al. in [32] suggest that the speed estimation error from EKF is approximately 2%. P. P. Acarnley assert that this application is limited when a low accurate is needed such as some general-purpose applications, not suitable for high-performance servo drives [32]. However, in our results, the error is less than 0.015 rad/s and represent only 0.0067% of the final value as it is depicting by Figure 6.

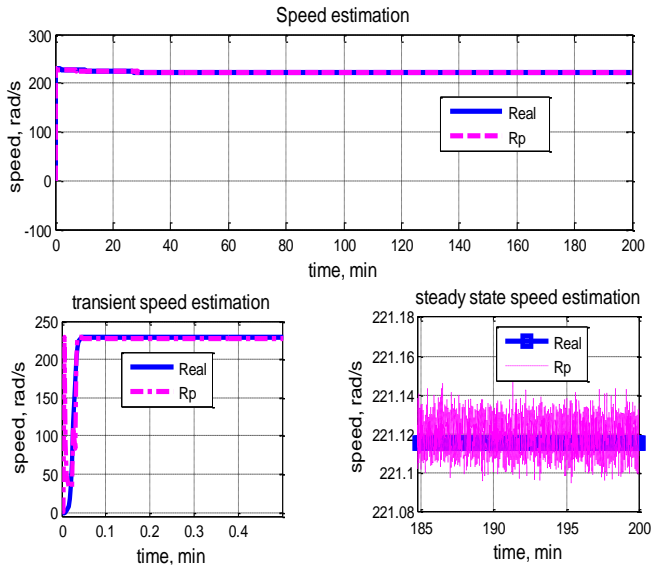

**Figure 3.** Estimated and simulated speed

Figure 4 presents the estimated armature temperature of a DC machine based on NN. As shown in Figure 4, the estimated temperature reaches 77 °C, and the model output nearby in the vicinity of 80 °C, while the steady state estimated error is less than 3 °C as can be seen from Figs. 6. However, Nestler et al. in [28] use a Luenberger's observer and it was shown that the estimated winding temperature error is important, and the results offered by Acarnley et al. in [32] concentrated in the same context and suggest that the temperature estimation error from EKF is 3 °C is approximately 3.75%.

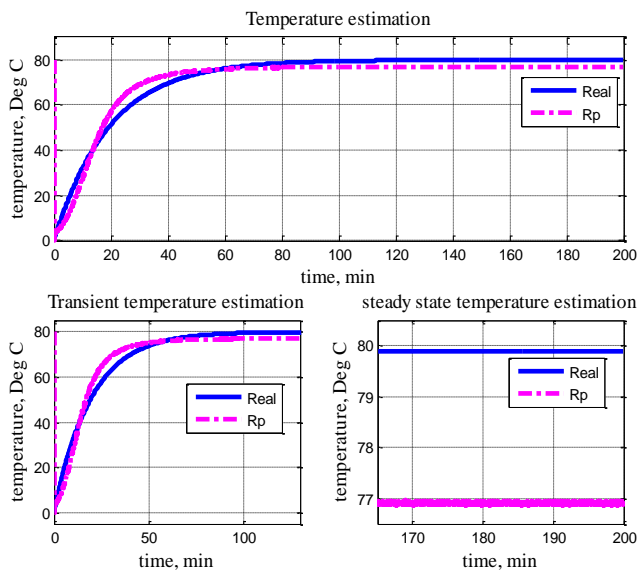

**Figure 4.** Estimated and simulated armature temperature.

Figure 5 depicts the estimate resistance by NN and the model response, from this figure, it can be seen that the resistance has the same curvature as the armature temperature, wherein the steady state the estimated resistance reached almost 4,56 Ω less than 0.04 Ω of simulated resistance, practically, this difference is negligible quantity and represents only 0.9 % of the final value, this results in this paper are more precise than the Zhang et al. results presented in [30], also this results are in agreement with the Karanayil et al. results presented in [91], where the errors of estimation of the rotor and stator resistances is 0.3% and 5% respectively.

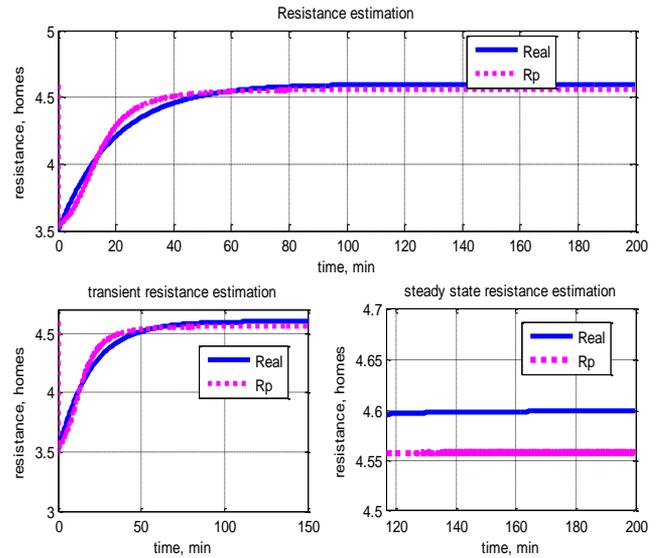

**Figure 5.** Estimated and simulated armature resistance

Figure 6 shows the estimation errors of speed, temperature and resistance, and their percentage in relation to their nominal value, this figure shows more clearly the perfect agreement between the model outputs and the intelligent sensor outputs.

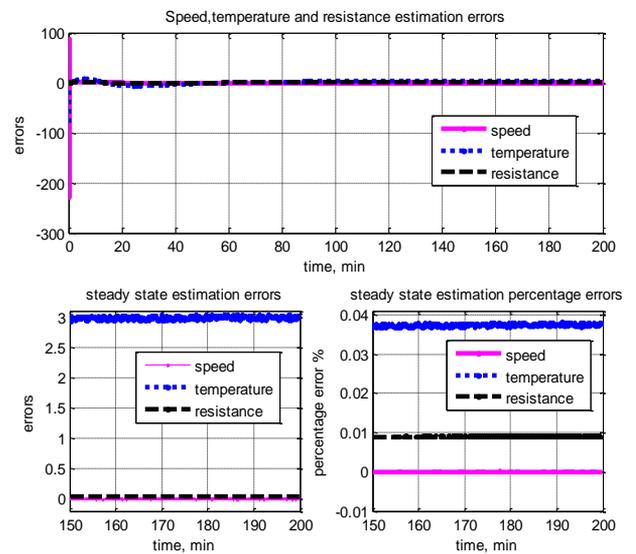

**Figure 6**. Speed, temperature and resistance estimation errors.

The following Table 2 summarizes the simulation errors in the steady state for all the estimated quantities by the ANN of

CFNN type based on a resilient backpropagation learning algorithm.

**Table 2.** Synopsis the estimation errors in steady-state

|  | Absolute error | Relative error |
|---|---|---|
| **Speed** | 0.015 rad/s | 0.0067% |
| **Temperature** | 3 $^0$C | 3.75% |
| **Resistance** | 0.04 Ω | 0.9% |

**Table.** Summary of the

## 5. CONCLUSIONS

A sensorless speed and armature winding quantity estimator is proposed for brushed DC machines based on CFNN trained by RBP. The proposed estimator includes a sensorless speed estimation, average armature temperature and resistance estimations based only on the voltage and the current measurements. The estimated speed and temperature eliminate the need for speed measurements and the need for the thermal sensor. In addition, the estimated temperature solves the problems of obtaining the thermal information from the rotating armature. Furthermore, the estimated resistance can be used to improve the accuracy of the control algorithms which are affected by an increase in resistance as a function of temperature. The good agreement between the model and the intelligent estimator demonstrates the efficiency of the proposed approach.

## ACKNOWLEDGMENT

This work was supported in part by: Laboratory Automation Systems (LAS) in the Electrical Engineering Department, Ferhat Abbas Setif1 University under the grant : A01 L07 UN1901 2019 0002, in other part by LGEER Laboratory, Hassiba Benbouali University, Chlef, Algeria, under the tutelage of the Algerian ministry of research and High education.

**NOMENCLATURE**

| | |
|---|---|
| b | viscous friction constant, N. m. s |
| E | sum squared error |
| H | thermal capacity, kJ. K$^{-1}$ |
| i | current, |
| J | total inertia, kg.m$^2$ |
| *K* | thermal transfer coefficients, W. K$^{-1}$ |
| k | loss constant, W. rad$^{-2}$. s$^2$ |
| *KS* | thermal transfer coefficients with speed, s. rad$^{-1}$ |
| ke | torque constant, V. rad$^{-1}$. s$^1$ |
| *l* | Inductance, H |
| net | weighted sum of the input unit |
| P | power, W |
| R | resistance, Ω |
| T | torque, N. m |
| V | voltage, V |
| y | network outputs |

**Greek symbols**

| | |
|---|---|
| α | temperature coefficient of resistance, K$^{-1}$ |
| θ | temperature above ambient, K |
| ω | armature speed, rad. s$^{-1}$ |
| Δ | weight step |
| η | factor |

**Subscripts**

| | |
|---|---|
| a | armature |
| a0 | ambient temperature |
| cu | Copper |
| d | desired |
| ir | iron |
| lo | losses |
| s | speed |
| max | maximum |
| min | minimum |
| 0 | Zero speed |
| − | decrease |
| + | increase |
| l | load |